\documentclass[aps,prl,10pt,twocolumn,superscriptaddress]{revtex4}
\newcommand{\bcso}{BaCuSi$_2$O$_6$}
\usepackage[latin1]{inputenc}
\usepackage{graphicx,epsfig,epstopdf,times}
\usepackage{dcolumn,color}
\usepackage{bm}
\usepackage{latexsym}
\usepackage{amssymb}
\usepackage{amsfonts}
\usepackage{amsmath}
\usepackage{graphicx}
\def\be{\begin{equation}}
\def\ee{\end{equation}}
\def\bc{\begin{center}}
\def\ec{\end{center}}
\newcommand{\bea}{\begin{eqnarray}}
\newcommand{\eea}{\end{eqnarray}}

\def\vec{\mathbf}

\newcommand{\veck}{{\vec k}}
\newcommand{\vecr}{{\bf r}}
\newcommand{\dagga}{{\phantom{\dagger}}}

\usepackage[colorlinks,bookmarks=false,citecolor=blue,linkcolor=red,urlcolor=blue]{hyperref}
\begin{document}
\title{Condensate-free superfluid induced by frustrated proximity effect}
\author{Nicolas Laflorencie}
\affiliation{Laboratoire de Physique des Solides, Universit\'e Paris-Sud, UMR-8502 CNRS, 91405 Orsay, France}
\author{Fr\'ed\'eric Mila}
\affiliation{Institut de Th\'eorie des Ph\'enom\`enes Physiques, \'Ecole Polytechnique F\'ed\'erale de Lausanne, CH-1015 Lausanne, Switzerland}
\begin{abstract}
Since the discovery of superfluidity in $^4$He and Landau's phenomenological theory, the relationship between Bose condensation and superfluidity has been intensely debated. $^4$He is known by now to be both superfluid and condensed at low temperature, and more generally, in dimension D$\ge$2, all superfluid bosonic models realized in experiments
are condensed in their ground-state, the most recent example being provided by ultracold bosonic atoms trapped in an optical lattice. In this paper, it is shown that a 2D gas of bosons which is not condensed at $T=0$ can be achieved by populating a layer through a {\it frustrated proximity effect} from a superfluid reservoir. This condensate-free bosonic fluid is further shown to be a superfluid with incommensurate correlations. \end{abstract}
\maketitle
The low energy properties of many condensed-matter systems are most naturally described
in terms of bosonic particles~\cite{Nozieres-book,Leggett-book}. This is of course the case of $^4$He~\cite{Matsubara56,Krauth91} and of cold bosonic
atoms loaded in optical lattices~\cite{Bloch08}, but effective bosonic models have also provided an accurate
description of several systems such as superconducting thin films~\cite{Jaeger89}, Josephson junction arrays~\cite{Fazio01}, or
quantum magnets in a field~\cite{Giamarchi08}.
At low enough temperature, most bosonic systems exhibit two remarkable quantum effects: Bose-Einstein condensation (BEC) and superfluidity. These two manifestations of quantum coherence often come together,
but this need not be the case: for instance, free bosons condense but are not superfluid,
while in principle superfluidity does not require a condensate, as first pointed out by Landau in his phenomenological
theory of superfluid $^4$He~\cite{Landau41}. However, to destroy the condensate while keeping a system superfluid
at $T=0$ in
realistic situations turns out to be tricky.
It is well-known~\cite{book-BEC} that interactions induce quantum fluctuations and generically tend to deplete the
condensate in the ground-state (GS), but they usually do not empty it completely. For instance, in $^4$He, despite
the very strong short-range repulsion, a condensate fraction of $\rho_0\sim 7\%$ of the total density remains~\cite{Ceperley95,Glyde00}. Generally speaking, interactions are known to deplete completely the $T=0$ condensate only for 1D geometries because of diverging fluctuations and for long-ranged interactions in the 2D Bose Coulomb liquid model with $\ln (r)$ interactions~\cite{Magro94},
a model with no experimental implementation so far. Increasing further interactions will ultimately induce a crystallized phase, which is of course neither condensed nor superfluid.
A realistic example for D$\ge 2$ where superfluidity could occur independently of BEC in the GS of a bosonic system is clearly missing. Here we explore a new route, using geometrical frustration to achieve a realistic condensate-free superfluid in D$=2$.
\begin{figure}
\begin{center}
\includegraphics[width=5cm,clip]{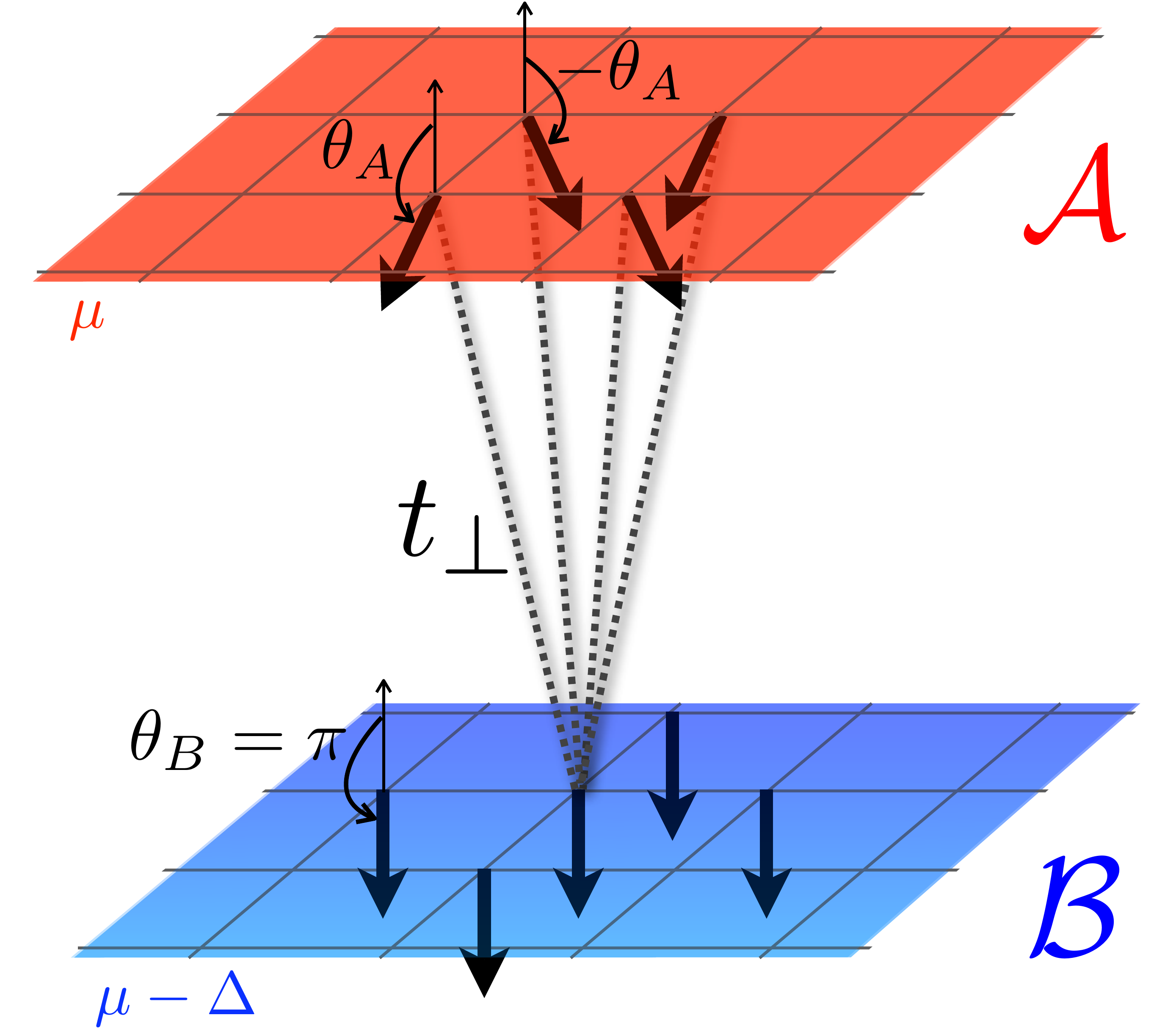}
\end{center}
\caption{Schematic picture for the frustrated bilayer made of two non-equivalent planes $\cal A$ and $\cal B$ which are coupled by a frustrated hopping term $t_{\perp}$. The thick arrows depict the spin-1/2 degrees of freedom [equivalent to hard-core bosons, see the text], which have their transverse component AF ordered [bosons condensed at ($\pi,\pi$)] in $\cal A$ layer while geometric frustration prevent AF order in $\cal B$ layer which, at the classical level, remain fully polarised [insulator for the bosons].}
\label{fig:schem}
\end{figure}
In the context of quantum magnetism, frustration has long been identified
as a possible mechanism to produce spin liquid phases~\cite{Anderson73,Balents10}, but it has only been realized recently that for bosonic models it could also induce new states of matter such as the supersolid state~\cite{Wessel05,Balibar10} or the putative Bose metal phase~\cite{Das99,Paramekanti02}.

In this Letter, we show that the GS of a 2D bosonic fluid created by a frustrated tunneling from a reservoir (as depicted in Fig.~\ref{fig:schem}) can be superfluid and not condensed. This conclusion is quite general and only relies on the frustrated character of the coupling. It thus applies in principle to a variety of systems such as optical lattices loaded with bosonic atoms or frustrated dimers in a strong magnetic field like \bcso, as we will discuss at the end of the paper.
The starting point is a simple 2D model of hard-core bosons on a square lattice:
\begin{equation}
{\cal{H}}=t\sum_{{\bf r},\tau_{\parallel}}(a^\dagger_{\bf r} a_{{\bf r}+\tau_{\parallel}}^{\dagga}+{\rm h.c.})-\mu \sum_{\bf r} n_{\bf r},
\label{eq:hcb}
\end{equation}
where $\tau_{\parallel}$ runs over the two basis vectors of the square lattice.
The hard-core constraint is essential to get superfluidity because it yields a linear excitation spectrum, which implies a finite critical velocity~\cite{Bernardet02}. On the other hand, such an infinite local repulsion is expected to deplete the condensate with respect to free bosons. To quantify these
effects, it is expedient to map the problem onto a spin model using the Matsubara-Matsuda
transformation~\cite{Matsubara56}: $S_{\bf r}^z=n_{\bf r}-1/2$, $S_{\bf r}^+=a_{\bf r}^\dagger$, $S_{\bf r}^-=a_{\bf r}$, which, up to a constant,
leads to an antiferromagnetic (AF) XY model in a transverse field.
\begin{equation}
{\cal H}_{\rm spin}=J\sum_{{\bf r},\tau_{\parallel}} (S_{\bf r}^xS_{{\bf r}+\tau_{\parallel}}^x+S_{\bf r}^yS_{{\bf r}+\tau_{\parallel}}^y)-H\sum_{\bf r} S_{\bf r}^z,
\end{equation}
with $J=2t$ and $H=\mu$. The basic physics already appears at the
classical level. For large negative $\mu$, the system is empty (all the spins point
down: $\langle S^{z}_{\bf r}\rangle=-1/2$). At $\mu=\mu_c\equiv -4t$ ($H=H_c\equiv -2J$), the bottom of the quadratic single particle dispersion band $\varepsilon({\bf k})=2t\left(\cos k_x +\cos k_y\right)-\mu$ vanishes at ${\bf k}_0=(\pi,\pi)$. This momentum defines the condensate mode in which bosons start to accumulate. The condensate density, defined on a lattice with $N$ sites by
\begin{equation}
\rho_{0}\equiv\frac{1}{N}\langle a^\dagger_{\bf k_0} a^{\dagga}_{\bf k_0} \rangle = \frac{1}{N^2} \sum_{{\bf r},{\bf r'}} {\rm{e}}^{i{\bf{k}}_0\left({\bf r}-{\bf r'}\right)}\langle S_{\bf r}^+S_{\bf r'}^- \rangle,
\end{equation}
is, in the magnetic language, equal to the AF order parameter in the plane perpendicular to the field.
The superfluid density $\rho_{s}$ is proportional to the stiffness (or helicity modulus) $\Upsilon$ of the system, i.e. the second derivative of the energy with
respect to a twist angle enforced at the boundaries~\cite{Fisher73}. At the classical level, the AF coupling between the XY components of
neighboring spins induces a planar long-range N\'eel order, which in the bosonic language leads to non-zero condensate and superfluid densities given by
$\rho_0=\rho_{s}=\rho(1-\rho)$,
where $\rho=1/2+\mu/8t$ is the bosonic density~\cite{Bernardet02}. So, at this level of approximation, the condensate
and superfluid densities are strictly equal.
Quantum fluctuations slightly reduce the condensate and
increase the stiffness~\cite{Bernardet02}, but in the simple model of Eq.(1),
the condensate persists at all densities. So, a single layer of hard-core bosons is a good prototype of
a bosonic system in D$\ge 2$, with both a condensate and a superfluid density at $T=0$.\\
\\
\noindent{\it{Bosonic bilayer}}---
We want to populate a layer through quantum tunneling from a superfluid reservoir. Therefore, we consider two copies $\cal A$ and $\cal B$ of the model~(\ref{eq:hcb})
\begin{eqnarray}
{\cal H}^{\cal{A}}&=&t\sum_{{\bf r},\tau_{\parallel}}(a^\dagger_{\bf r} a_{{\bf r}+\tau_{\parallel}}^{\dagga}+{\rm h.c.})-\mu \sum_{{\bf r}} n_{\bf r}^{\cal A}\label{eq:HbosonsA}\\
{\cal H}^{\cal{B}}&=&
t\sum_{{\bf r},\tau_{\parallel}}(b^\dagger_{\bf r} b_{{\bf r}+\tau_{\parallel}}^{\dagga}+{\rm h.c.})-(\mu-\Delta) \sum_{{\bf r}} n_{\bf r}^{\cal B},
\label{eq:HbosonsB}
\end{eqnarray}
where $\Delta>0$ is an energy barrier.
The critical chemical potentials now take different values:
on the $\cal A$ layer, $\mu_c=-4t$, whereas on the $\cal B$ layer, $\mu_c=-4t+\Delta$,
and if $0<\mu-\mu_c<\Delta$, the $\cal A$ layer has a finite density while the $\cal B$ layer is empty. Let us now consider, in such a regime, the effect of a frustrated coupling between the layers
\begin{equation}
H^{\cal AB}=t_{\perp}\sum_{\vecr,\tau_{\perp}} (a_{\vecr+\tau_{\perp}}^\dagger b_{\vecr}^{\dagga}+b_{\vecr}^\dagger a_{\vecr+\tau_{\perp}}^{\dagga}),
\label{interlayer_boson}
\end{equation}
where $\tau_{\perp}$ runs over the four vectors coupling one site of the ${\cal B}$ layer to its four nearest neighbours
in the ${\cal A}$ layer. As depicted in Fig.~\ref{fig:schem}, in the magnetic representation it is easy to see that such a transverse frustrated coupling leads, 
at the classical level, to a vanishing local $xy$ field at each site of the ${\cal B}$ layer, and therefore prevents bosons from $\cal A$ to tunnel into $\cal B$.  
Such a classical decoupling has to be contrasted with the case of a direct unfrustrated tunneling of the form
$t_{\perp}\sum_{\bf r} (a^\dagger_{\bf r} b_{\bf r}^{\dagga}+b^\dagger_{\bf r} a_{\bf r}^{\dagga})$ for which a local field is immediately induced in the $\cal B$ layer, leading to a finite density $\rho^{\cal B}$.

\begin{figure}
\begin{center}
\includegraphics[width=0.9\columnwidth,clip]{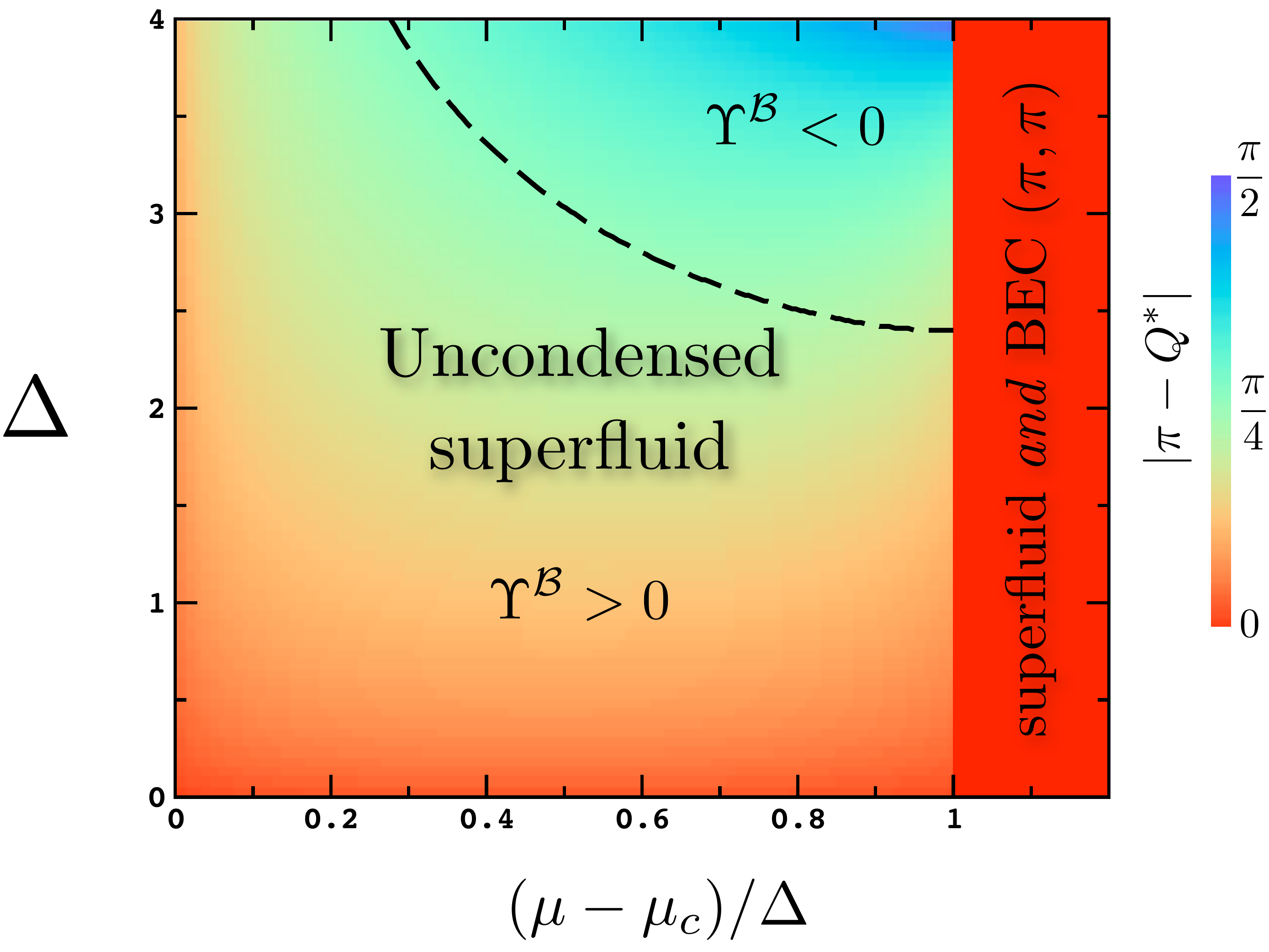}
\end{center}
\caption{Phase diagram of the bosonic gas of the $\cal B$ layer for the frustrated bilayer system, with the normalised chemical potential $(\mu-\mu_c)/\Delta$ on the x-axis and the energy barrier $\Delta$ between layers $\cal A$ and $\cal B$ on the y-axis. Colours indicate the values of the incommensurate vector $Q^*$. For $\mu-\mu_c>\Delta$ $Q^*=k_0=\pi$, superfluidity and BEC occurs together (red region) whereas for $\mu-\mu_c<\Delta$, there is no true BEC but a finite superfluid density $\rho_{s}^{\cal B}$ while the stiffness $\Upsilon^{\cal B}$, positive for not too large $\Delta$ can change sign and become negative for larger $\Delta$ (see text).}
\label{fig:PHDG}
\end{figure}

Coming back to the frustrated case of Eq.~(\ref{interlayer_boson}), since the local field vanishes, the ${\cal A}$ layer has no influence at the classical level on the ${\cal B}$ layer,
which remains empty.
This cannot be true however when {{many-body effects}} and quantum fluctuations
are included. Indeed, in the bosonic language,
a wave-function with particles only in the $\cal A$ layer is not an eigenstate when one includes the
interlayer hopping of Eq.~\ref{interlayer_boson}.
So particles have to be present in the $\cal B$ layer. This does not mean however that there is a condensate. Indeed, according to field theory, a direct (linear) coupling between condensate order parameters on $\cal A$ and $\cal B$ is incompatible with the symmetry of the lattice~\cite{Rosch07a}. In our case of non-equivalent layers, this implies that a condensate in the $\cal B$ layer will not develop immediately when a condensate appears in the $\cal A$ layer. As we shall see, treating quantum fluctuations at the level of linear spin-wave (LSW) theory 
indeed leads to a small bosonic population in the ${\cal B}$ layer which is not condensed. We now explore the very peculiar properties of this bosonic gas.\\
\\
{\it{Linear spin-wave corrections---}}
To treat quantum fluctuations, we start from the spin representation and perform a Holstein-Primakoff
transformation~\cite{Holstein40} after a local rotation of the spins. In the classical GS, the spins in the
$\cal A$ layer make an angle $\pm\theta_{\cal A}=\pm\arccos(-\mu/\mu_c)$ with the $\hat z$ direction, while in
the $\cal B$ layer the spins point opposite to $\hat z$, as sketched in Fig.~\ref{fig:schem}. To avoid confusion with the original bosonic operators,
the Holstein-Primakoff bosons on layers $\cal A$ and $\cal B$ are denoted by $\tilde a_i^\dagger,\tilde a_i$ and
$\tilde b_i^\dagger,\tilde b_i$. The resulting Hamiltonian is quadratic, and can be diagonalized
by a Fourier transformation followed by a Bogoliubov transformation, which leads to the diagonal Hamiltonian:
\be
{\cal H}_{\rm LSW}=\sum_{\bf k}\left(\Omega^{\alpha}_{\bf k}\alpha_{\bf k}^{{\dagger}}\alpha_{\bf k}^{\vphantom{\dagger}}+\Omega^{\beta}_{\bf k}\beta_{\bf k}^{{\dagger}}\beta_{\bf k}^{\vphantom{\dagger}}\right)+{\rm{const.}},
\label{eq:LSW}
\ee
where the Bogoliubov operators $\alpha_{\bf k}^{{\dagger}}$, $\beta_{\bf k}^{{\dagger}}$ are linear combinations of the Hostein-Primakoff operators $\tilde a_{\bf r}^\dagger,\tilde a_{\bf r}$ and
$\tilde b_{\bf r}^\dagger,\tilde b_{\bf r}$.
The new GS is now the vacuum of the Bogoliubov quasi-particles: $\langle \alpha_{\bf k}^{{\dagger}}\alpha_{\bf k}^{\vphantom{\dagger}}\rangle=\langle \beta_{\bf k}^{{\dagger}}\beta_{\bf k}^{\vphantom{\dagger}}\rangle=0$, {{and inverting the relation between the Bogoliubov operators and the Holstein-Primakoff operators gives access to the
GS properties of the system.}}
\begin{figure}
\begin{center}
\includegraphics[width=0.8\columnwidth,clip]{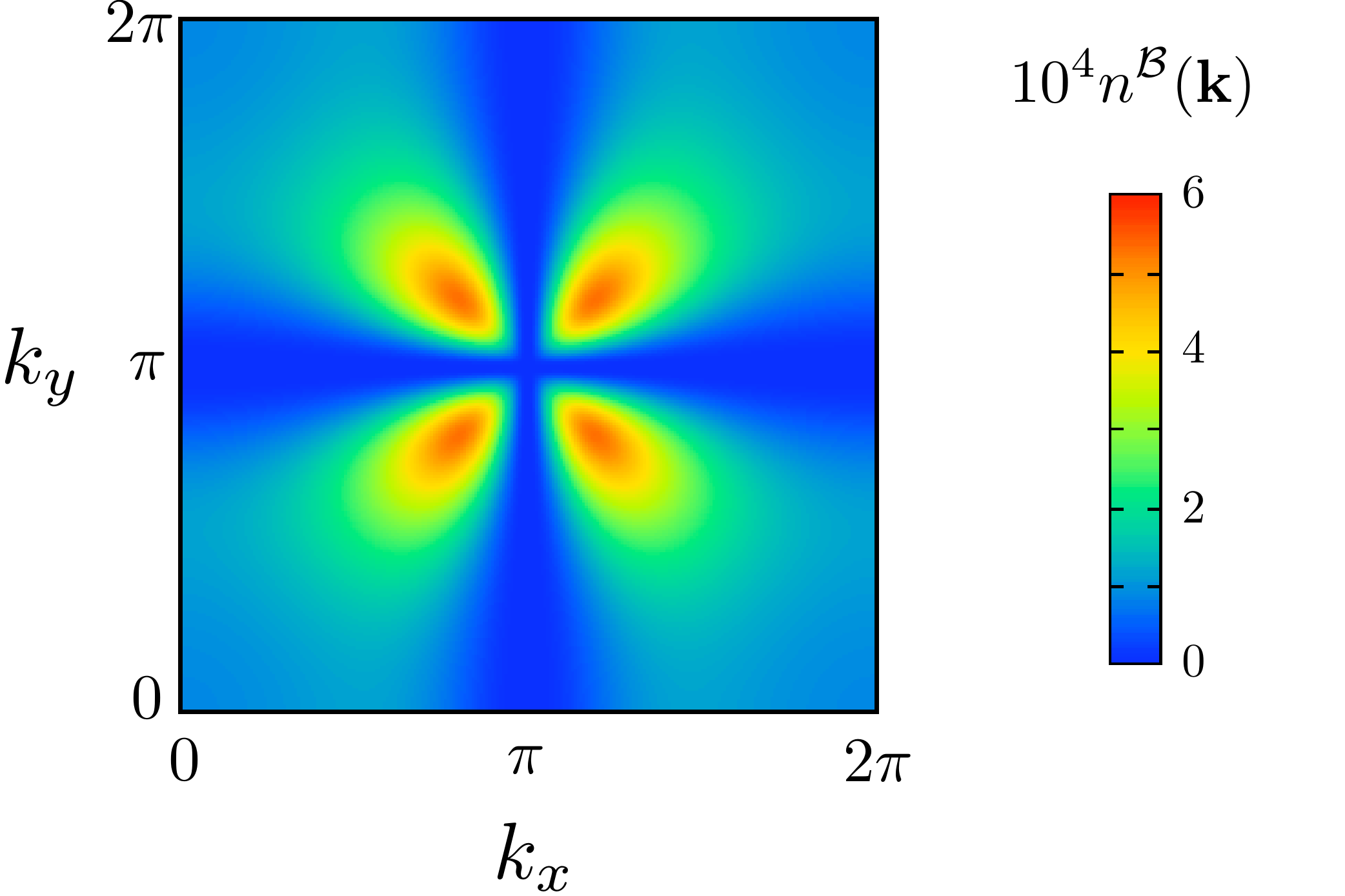}
\end{center}
\caption{Momentum distribution of the LSW modes on the $\cal B$ layer: $n^{\cal B}(k_x,k_y)$ for $t=1$, $t_{\perp}=0.5$, $\Delta=1$, and $\mu-\mu_c=\Delta/2$}
\label{fig:nk3d}
\end{figure}
%
%
The $\cal A$ layer is almost unaffected by the presence of the $\cal B$ layer: the density
grows linearly above $\mu_c$ (strictly speaking, additional small logarithmic corrections are expected in 2D), as in an isolated layer, and condensate and superfluid densities
are almost equal. The physical properties of the $\cal B$ plane are very different however from those
of an isolated layer.
To describe them more precisely, we consider the total density of bosons
$\rho^{\cal B}_{\rm tot}\equiv \frac{1}{N}\sum_{\vec k} \langle b^\dagger_{\vec k} b_{\vec k}^{\dagga} \rangle
$, the condensate density
$
\rho^{\cal B}_{0}\equiv\frac{1}{N}\langle b^\dagger_{\vec k_0}  b_{\vec k_0}^{\dagga} \rangle
$, and the superfluid stiffness $\Upsilon^{\cal B}$, defined by introducing a small twist angle $\varphi$ along one direction of the ${\cal B}$ plane only. The hopping amplitude $t$ in ${\cal H}^{\cal{B}}$ [Eq.~(\ref{eq:HbosonsB})] is replaced by $t{\rm{e}}^{{\rm{i}}\varphi}$ along this direction, and the superfluid stiffness is given by $\Upsilon^{\cal B}=\left({\partial^2 e}/{\partial \varphi^2}\right)_{\varphi=0}$, $e$ being the energy density of the system. This leads to the expression:
$\Upsilon^{\cal B}=-\frac{t}{N}\sum_{\bf k}\langle b^{\dagger}_{\bf k}b_{\bf k}^{\vphantom \dagger}\rangle\left(\cos k_x +\cos k_y\right)$.
A key quantity here is the momentum distribution $n^{\cal B}({\bf k})=\langle b^{\dagger}_{\bf k}b_{\bf k}^{\vphantom \dagger}\rangle$ 
that we show in Fig.~\ref{fig:nk3d} for a representative set of parameters. It displays 4 incommensurate maxima away from the BEC point ${\bf k}_0$ at ${\bf Q}^*=(\pi\pm q^*,\pi\pm q^*)$ whose position in the Brillouin
zone (BZ) depends on $\Delta$ and $\mu$. Perhaps more importantly, the momentum distribution vanishes when ${\bf k}\to{\bf k}_0$ as $n^{\cal B}(\bf k)\sim \|\bf k-{\bf k}_{0}\|^3$ as we discuss below. This remarkable behaviour results in the following properties for the bosonic fluid  in $\cal B$: (i) the total density $\rho_{\rm tot}^{\cal B}$ is non zero as soon as $\mu>\mu_c$~\cite{Note1}; (ii) the condensate mode remains empty: quantum fluctuations do not change the classical result $\langle b^\dagger_{\vec k_0} b^{\vphantom \dagger}_{\vec k_0}\rangle=0$ in the GS and the distribution $n^{\cal B}({\bf k})$ is never singular, which signals the absence of condensate at any vector $\bf k$; (iii) there is a finite superfluid stiffness which is strongly influenced by the location in the BZ of the incommensurate vector ${\bf Q^*}$ where the response is maximal. It is positive and leads to a superfluid density $\rho^{\cal B}_{s}=\Upsilon^{\cal B}/2t$ of the same order as
the total density if ${\bf Q^*}$ is close to the BEC point $\veck_0$. However, it strongly decreases when ${\bf Q^*}$ shifts away from
$\veck_0$, and it eventually changes sign for large enough $\Delta$ when the maximum gets far enough from $\veck_0$, which signals an instability towards spontaneous currents, as discussed some time ago in the context of dirty superconductors~\cite{Spivak91}.\\
%
%
\\
\noindent{\it{Filtering and frustrated proximity effects}}---
The basic physical mechanism behind the absence of a condensate and the incommensurate fluctuations
in the $\cal B$-layer is actually most simply understood as a filtering mechanism due to frustration.
To get an intuitive picture of this effect, it is useful to look at the single particle dispersion
\bea
\varepsilon(k_x,k_y)&=&-2t\left(\cos k_x +\cos k_y\right)-\mu+
\frac{\Delta}{2}\\
&\pm&\sqrt{\frac{\Delta^2}{4}+4t_{\perp}^2\cos^2\left(\frac{k_x}{2}\right)
\cos^2\left(\frac{k_y}{2}\right)}\nonumber.
\eea
The inter-layer delocalization is governed by an effective hopping amplitude $t_{\perp}\cos \left(\frac{k_x}{2}\right)\cos\left(\frac{k_y}{2}\right)$ which vanishes exactly at the BEC point ${\vec k}_0$. Therefore, a single particle injected in the $\cal A$ layer with momentum ${\vec k}_0$ is confined to this
layer.
\begin{figure}
\begin{center}
\includegraphics[width=0.8\columnwidth,clip]{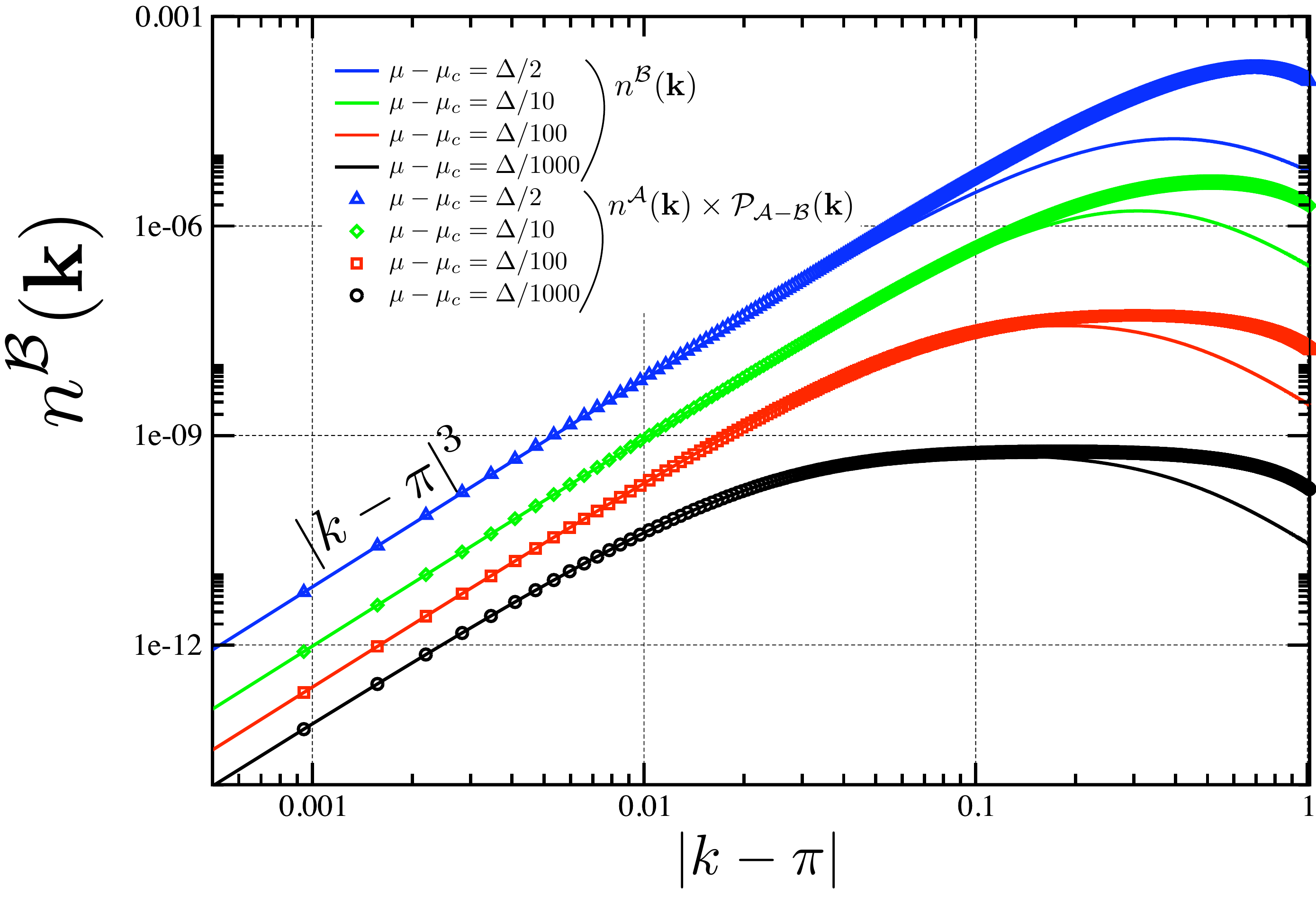}
\end{center}
\caption{$n^{\cal B}(k)$ along the line $k_x=k_y=k$ compared to $n^{\cal A}(k)\times {\cal P}_{\cal A-B}(k)$ discussed in the text. The parameters used are $t=1$, $t_{\perp}=0.1$, $\Delta=1$ at various distances from the critical point $\mu-\mu_c$.}
\label{fig:rhoB_Qstar}
\end{figure}
At second order in $t_\perp/\Delta$, the probability for a boson of momentum $\bf k$ to tunnel from $\cal A$ to $\cal B$ is given by
\be
{\cal{P}}_{\cal A - B}({\bf k})\propto\left[\frac{t_{\perp}}{\Delta}\cos\left(\frac{k_x}{2}\right)\cos\left(\frac{k_y}{2}\right)\right]^2.
\ee
As long as the density in the $\cal A$ layer is not too large, this single particle tunneling probability can be
used to estimate the occupation in $\cal B$, which, since all particles come from $\cal A$ through
tunnelling, is expected to be approximately given by:
\be
n^{\cal B}({\bf k}) \simeq n^{\cal A}({\bf k})\times {\cal{P}}_{\cal A - B}({\bf k}).
\ee
In the vicinity of the BEC mode $\veck_0$, the hopping probability vanishes very rapidly, as ${\cal{P}}_{\cal A-B}\sim \|\veck_0-{\bf k}\|^4$. Combined with the fact that the zero-mode fluctuations in the $\cal A$ layer diverge like $n^{\cal A}\sim 1/ \|\veck_0-{\bf k}\|$, this leads to $n^{\cal B}({\bf k})\sim \|\veck_0-{\bf k}\|^3$ for ${\bf k}\to \veck_0$, in agreement with the exact evaluation of $n^{\cal B}({\bf k})$ from the Bogoliubov transformation (see Fig.~\ref{fig:rhoB_Qstar}).
More interestingly, this ${\bf k}$-dependent tunneling mechanism provides a simple explanation of the maximal response observed in $\cal B$ at the incommensurate vector ${\bf Q}^*$, away from the condensate point $\veck_0$.
Indeed, there is a threshold vector $\kappa(\mu)\sim (\mu-\mu_c)^{1/4}$ such that for $\|\veck_0-{\bf k}\|>\kappa$ the occupation in $\cal A$ start to decay faster than $1/\|\veck_0-{\bf k}\|^4$, which leads to a maximal response at $|\pi-Q^*|\sim \kappa$. 
The nature and the properties of this unconventional quantum liquid are summarized in Fig.~\ref{fig:PHDG}.\\
\\
{\it{Experimental consequences---}} Since the filtering mechanism induced by a frustrated coupling between bosonic layers yields a condensate-free superfluid state with incommensurate correlations, it is legitimate to ask whether its experimental realization is possible or not. As discussed in more details in the supplementary material, two directions can be considered, namely ultracold bosonic atoms loaded in an optical lattice, and the quantum antiferromagnet \bcso~in a field. For the former example, while the harmonic trap will break the translational invariance and slightly perturb the perfect frustration between layers, we can extend the SW calculation to the inhomogeneous case (see supplementary material) to show that, in practice, it will not qualitatively change the predictions made for the homogeneous case. Concerning the frustrated magnet \bcso~in a strong external field, as also discussed in the supplementary material, it is a very interesting 3D realization of the frustrated layered model with two types of layers~\cite{Ruegg07,Kramer07,Laflorencie09}. While the formal analogy between superfluids and quantum magnets is limited, we predict that high field NMR and neutron experiments may detect the absence of field-induced triplets BEC and the presence of incommensurate correlations in half the layers of the quantum antiferromagnet \bcso.\\
\\
\noindent{\it{Conclusions}}---
We have shown in the context of a simple hard-core boson model of two coupled planes that frustration can influence
dramatically the proximity effect induced by a hopping term between the layers. If the frustration is such that
the condensate of one layer cannot tunnel to an otherwise empty layer, then the bosonic gas induced in this
layer by tunnelling has been shown to be superfluid but uncondensed. Beyond this simple model, this effect is
expected to be present whenever two bosonic systems, an occupied one and an empty one, are put into contact by a proximity
effect, provided the geometry is such that the condensate cannot hop from one system to the other.
This effect leads to non-trivial predictions for trapped bosons in an optical lattice where the presence of a quadratic potential does not qualitatively change the physics. Furthermore, the quantum antiferromagnet \bcso~in a field is also predicted to display a very rich physics for the triplet excitations in a window of $\sim 2$ Tesla above the critical field $H_c$.
We hope
that the present work will encourage experimental investigations in both directions, cold atoms and quantum magnets.\\
We thank C. Berthier, M. Horvatic and S. Kr\"amer for useful discussions on the properties of the Han purple
as revealed by NMR, and F. Alet and J.-N. Fuchs for a critical reading the manuscript. This work has been supported by the Swiss National Fund and by MaNEP.

\end{document}


\title{{{Supplementary material for "Condensate-free superfluid induced by frustrated proximity effect"}}}
\author{Nicolas Laflorencie}
\affiliation{Laboratoire de Physique des Solides, Universit\'e Paris-Sud, UMR-8502 CNRS, 91405 Orsay, France}
\author{Fr\'ed\'eric Mila}
\affiliation{Institut de Th\'eorie des Ph\'enom\`enes Physiques, \'Ecole Polytechnique F\'ed\'erale de Lausanne, CH-1015 Lausanne, Switzerland}
\maketitle
\section{Trapped bosons in optical lattices}
\begin{figure}
\begin{center}
\includegraphics[width=0.45\columnwidth,clip]{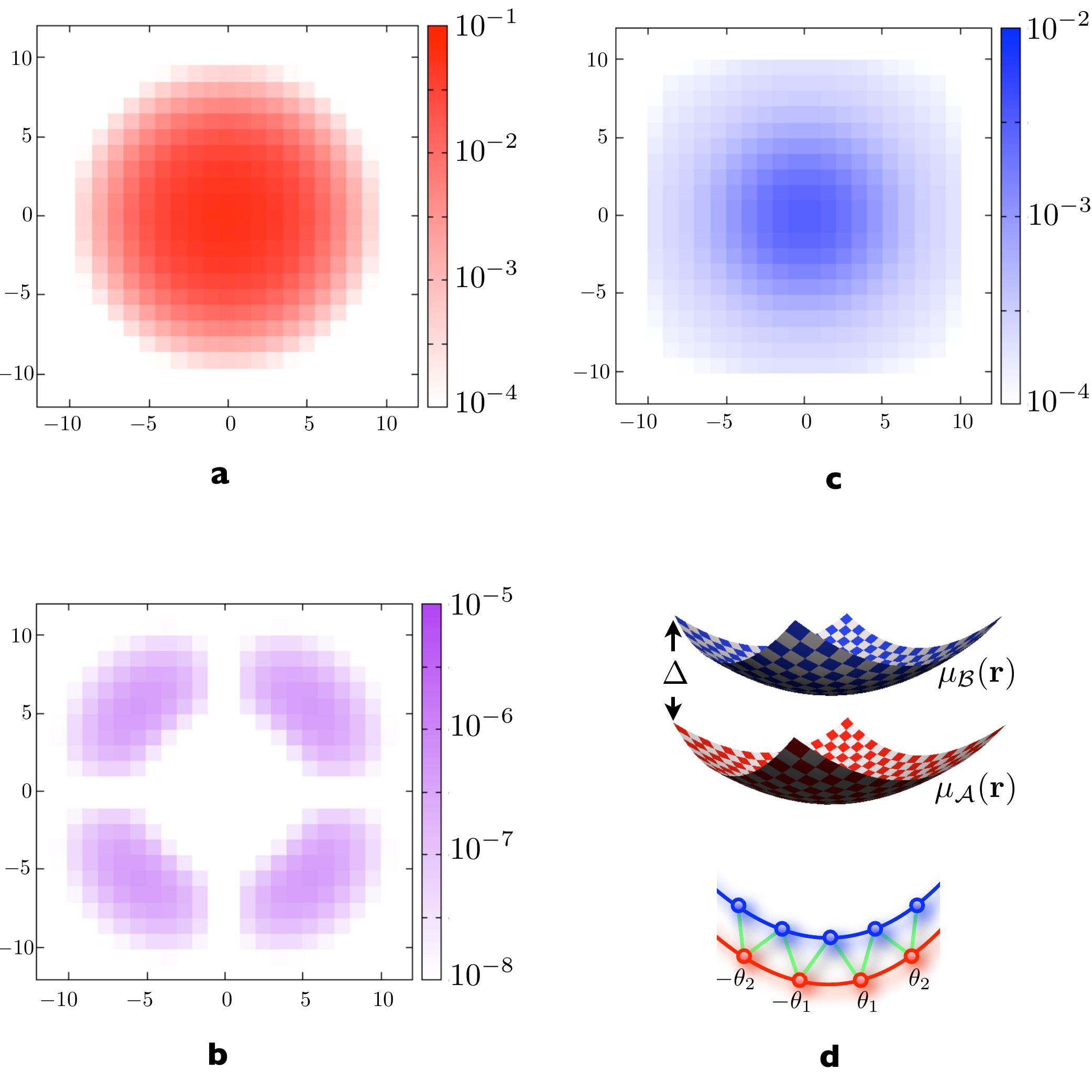}
\end{center}
\caption{{Real space density profiles for a trapped bilayer system of $24\times 24\times 2$ lattice sites, with a chemical potential $\mu=\mu_c+t/2$, a trapping potential $V_0=10^{-2}t$, a barrier $\Delta=4t$, and an interlayer hopping $t_{\perp}=0.5t$. Local densities at the classical approximation shown for layers $\cal A$ (a) and ${\cal B}$ (b). Quantum fluctuations induce a much larger population in $\cal B$ (c).
(d) Top: schematic picture for both layers $\cal A$ and $\cal B$ in a harmonic trap and separated by a barrier $\Delta$. Down: 1D representation of the frustrated setup which gradually loose perfect frustration away from the trap center.}}
\label{fig:density_trap}
\end{figure}

Experiments using ultra-cold gases have recently shown their ability to reproduce strongly correlated many-body physics~\cite{Bloch08}. In particular, when loaded in a D-dimensional optical lattice, trapped bosons may display quantum phase transitions such as the celebrated BEC-Superfluid to Mott-Insulator transition~\cite{Greiner02}. Another remarkable achievement concerns the interesting Tonks-Girardeau regime analog to hard-core bosons~\cite{Paredes04}. Recently, it has also been shown that artificial gauge fields~\cite{Lin09,Dalibard10} can be created in such systems, allowing to achieve positive hopping terms between nearest neighbors $-t\exp({\rm i}\pi)$, a necessary condition to frustrate the condensate mode in our case.
As compared to archetypal solid-state uniform bulk systems, the price to pay with optical lattices is double: finite system size (typically $10^{4}$ lattice sites) and a trapping potential which breaks the translational invariance.
The presence of a harmonic trap leads to a site-dependent chemical potential
\be
\label{eq:mur}
\mu({\bf r})=\left\{
\begin{array}{lcc}
\mu-V_0 r^2&{\rm{~if~}}& {\bf r}\in {\cal A}\\
\mu-V_0r^2-\Delta&{\rm{~if~}}& {\bf r}\in {\cal B}.
\end{array}
\right.
\ee
Usually, a method known as local density approximation is used to deal with inhomogeneous potentials~\cite{LDA}. However, this is not so simple here since frustration turns out to be quite sensitive to the inhomogeneities in the local densities. Indeed, contrary to the translationally invariant case where in the magnetic language a uniform antiferromagnetic XY order on the $\cal A$ layer yields a zero local mean field at each site of the $\cal B$ layer, the classical spin vectors ${\bf S}=S(\sin\theta,0,\cos\theta)$ have angles which are now smoothly site-dependent $\theta_{\cal A}({\bf r})$ within the trap. As a consequence, $\sum_{{\rm {pl.}}}\theta_{\cal A}({\bf r})$ (pl.=plaquette) will not necessarily vanish, as illustrated in Fig.~\ref{fig:density_trap} (d) in a 1D representation.

{{The classical solution is obtained by a global numerical minimization of the full set of local angles $\theta_{\cal A/B}({\bf r})$ on both layers. Unlike the uniform case ($V_0=0$), here an extremely small but finite population can hop from $\cal A$ to $\cal B$ already at the classical level, 
as visible in Fig.~\ref{fig:density_trap} (b). Nevertheless, this classical population has a peculiar spatial distribution due to the trade-off between two competing
effects. On one hand, in order to satisfy the local energetic constraints imposed by the chemical potential, particles would tend to be maximally present at the center of the trap with a density which decreases away for the trap center, as clearly observed for the $\cal A$ layer in Fig.~\ref{fig:density_trap} (a). On the other hand, the local constraint imposed by the interlayer frustration decreases with $r$.
At the very center, and along the axes of the square lattice, the symmetry imposes a perfect frustration $\sum_{{\rm {pl.}}}\theta_{\cal A}({\bf r})=0$ which gradually disappears away from these lines. These two competing effects give rise to a distribution in the $\cal B$ layer which is localized away from the center,
the square lattice axes, and the edge of the trap, as shown in Fig.~\ref{fig:density_trap} (b). Remarkably, the maximal density in $\cal B$ is 5 orders of magnitude below the maximal density in the $\cal A$ layer. This shows that the frustration effect, while formally unperfect, remains in practice very efficient, especially close to the center of the trap. Nevertheless the nature of this highly diluted bosonic population which has tunneled from $\cal A $ to $\cal B$ is conventional, i.e. condensed and superfluid but with a tiny density, almost experimentally undetectable.

As previously done for the uniform case, we investigate quantum fluctuations on top of the classical GS using the linear spin-wave approximation. Due to the lack of translational invariance, we must study quantum fluctuations in real space, which leads to an eigenvalue problem of much larger dimension (twice the number of sites instead
of 4). The GS
energy $E_0(\{\mu(\bf r)\})$ for a given set of chemical potentials is obtained as the sum of the classical energy and of the zero point
fluctuations, and the local density at a site $\bf r$ is obtained computing $-\partial E_0/\partial \mu(\bf r)$ evaluated
numerically at the chemical potentials of Eq.~(\ref{eq:mur}). As visible in Fig.~\ref{fig:density_trap} (c), the bosonic fluid induced by quantum fluctuations completely dominates the physics of the $\cal B$ layer, in which the classically empty center of the trap is now maximally populated. Most importantly, this new bosonic population, 3 order of magnitude above the classical one, will display similar properties as in the uniform case, i.e. absence of Bose condensate, a finite superfluid density, and incommensurate correlations.
{\red{Newly developed detection methods in real space~\cite{Sherson10,Bakr10} will allow to probe the bosonic population close to the trap center. The incommensurate response at $\bf Q^*$ can be detected by measuring the momentum distribution $n^{\cal B}({\bf k})$ using standard time of flight experiments~\cite{Paredes04}.}}
\section{\bcso~at high magnetic field}
{\red{As already discussed in Ref.~\cite{Ruegg07,Kramer07,Laflorencie09}, \bcso~under a strong magnetic field is a good realization of a 3D arrangement of inequivalent  bosonic planes $\cal A$-$\cal B$ which display a fully frustrated transverse coupling, and a bare gap $\Delta\sim 3$ T.}}
Building on our understanding of the two-layer boson system studied above, it is straightforward to explore the properties of the exotic triplet fluid on ${\cal B}$ layers within a LSW approach of the full 3D model with realistic parameters for \bcso. In a magnetic field range of $\sim 2$ T above the critical field $H_c=23.4$ T, the ${\cal B}$ layers will not contribute a Bragg peak at $\veck_0=(\pi,\pi)$, but rather develop a maximum at a wave-vector $\bf Q^*$ which is expected to display an unusual power-law dependence on the distance to the critical field $H_c$ with an exponent $1/4$:
\be
\|{\bf Q}^*-{\bf k}_0\|\propto \left(H-H_c\right)^{1/4}.
\ee
These features are direct consequences of the absence of condensate, one of the main predictions of the theory
developed in this paper. They are in principle accessible to NMR and neutron scattering. In NMR, the absence
of condensate implies that the massive broadening of the line that accompanies the development of magnetic 
order will not take place. The incommensurate response can in principle be observed in the line shape.
This will be complicated by the presence of an incommensurate distortion in the Han purple~\cite{Samulon06}, but the two 
effects might be disentangled by studying the evolution of the line shape with the field since the 
incommensurate response predicted by the present theory is field dependent, while the incommensurate 
distortion is not expected to be. The absence of a Bragg peak and the incommensurate response may also
be detected by neutron scattering. 
Indeed, using pulsed field, the possibility to perform elastic neutron scattering up to 30 T has been very recently demonstrated~\cite{Matsuda09}, thus opening the route to probe the peculiar field-induced momentum distribution in \bcso.
Let us reemphasize however that the other prediction of our theory, the persistence of superfluidity
despite the loss of condensate does not apply to the Han purple because of residual anisotropies~\cite{Sebastian06b} that break
the rotational symmetry.